Zoran Ivanovski (Corresponding author)[1]
Toni Draganov Stojanovski[2]
Nadica Ivanovska[3]

# INTEREST RATE RISK OF BOND PRICES ON MACEDONIAN STOCK EXCHANGE - EMPIRICAL TEST OF THE DURATION, MODIFIED DURATION AND CONVEXITY AND BONDS VALUATION

**Abstract**

This article presents valuation of Treasury Bonds (T-Bonds) on Macedonian Stock Exchange (MSE) and empirical test of duration, modified duration and convexity of the T-bonds at MSE in order to determine sensitivity of bonds prices on interest rate changes. The main goal of this study is to determine how standard valuation models fit in case of T- Bonds that are traded on MSE and to verify whether they offer reliable results compared with average bonds prices on MSE. We test the sensitivity of T- Bonds on MSE on interest rate changes and determine that convexity is more accurate measure as approximation of bond prices changes than duration. Final conclusion is that T-Bonds traded at MSE are not sensitive on interest rate changes due to institutional investors' permanent higher demand and at the same time market limited offer of risk-free instruments.

[1] Faculty of Economics, University of Tourism and Management in Skopje, Partizanski odredi 99, 1000 Skopje, Republic of Macedonia, z.ivanovski@utms.edu.mk

[2] Faculty of Informatics, European University Republic of Macedonia, Bul. Kliment Ohridski 68, 1000 Skopje, Republic of Macedonia, toni.draganov.stojanovski@eurm.edu.mk

[3] Central Cooperative Bank, Arena Filip II, 1000 Skopje, Republic of Macedonia, nadica.ivanovska@ccbank.com.mk

## 1. Valuation of Treasury Bonds at Macedonian Stock Exchange

The price of any financial instrument is equal to the present value of the expected cash flow from the financial instrument (Damodaran, Applied Corporate Finance, 1999). In order to determine intrinsic value of the bond we need to estimate expected cash flows and appropriate required rate of return (yield). The expected cash flows are determined from bond characteristics or bond contract. The required rate of return (yield) reflects the yield for financial instruments with comparable risk, or alternative investments (Brealey, 2006). Bond as a debt instrument requires from the issuer (debtor or borrower) to repay to the lender/investor the amount borrowed (principal) plus interest over a specified period of time. A key feature of a bond is the nature of the issuer, which is usually divided in three groups: government, municipalities and corporations (domestic and foreign).

We are analyzing Treasury Bonds (T-Bonds) that are quoted and traded on Macedonian Stock Exchange (MSE). MSE was established in September 1995, but its real start was with the first ring of Stock-Exchange bells on 28 March 1996. First Treasury Bonds' quotation on MSE happened in 2000, when Ministry of Finance issued T- Bonds as compensation for "frozen" (old) foreign-exchange deposits in Macedonian commercial banks before the Republic of Macedonia gained independence from former Yugoslavia (Code: RM01). New types of T- Bonds (Bonds for Denationalization, Code: RMDEN) were issued on MSE in 2002. Starting from 2006, MSE regularly calculates Bond Price Index (OMB).

MSE's short-history strongly affects securities valuation, due to the relatively short time series and impossibility to calculate market premium. The limited numbers of securities that are quoted and traded on MSE as well as the low liquidity of the market are additional factors that have significant influence on the process of valuation.

The Old Foreign-Exchange Saving Bonds and Bonds for Denationalization are listed and traded on MSE. Bonds for Denationalization are most liquid, permanently and in significant amounts traded on MSE. These types of bonds are the focus of our research. Pursuant to the Law on Issuance of Bonds for Denationalization, Republic of Macedonia in a period of eleven years carried out one issue of Bonds for

Denationalization annually. The Government of the Republic of Macedonia every year makes a decision on the amount of the Bonds for Denationalization to be issued. First issue of Bonds for Denationalization was made in March, 2002, while last eleventh issue of this type of bonds was launched in March 2012. Total amount of issue per year vary between EUR 10 - 30 million, and depends on amounts of effective Government decisions for denationalization made for specific year for which bonds are given as a form of compensation. Bonds for Denationalization are registered securities, denominated in euro and unrestrictedly negotiable. Face value of the bond is EUR 1. Interest and portion of face value of the bonds fall due on June 1 every year, which means that they are amortization bonds (with annuity payment of interest and principal) with 10 years maturity. Following the adoption of the request for listing on the official market, the bonds are traded on the secondary market of the MSE. Trading with the bonds, listed on the MSE, is carried out on the basis of the market price. Payment upon executed purchase of the bonds on the secondary market is carried out at the price at which they have been traded on the Stock Exchange, including the accrued interest for the period from the last interest payment up to the transaction day.

### 1.1. Calculation of Risk-Free Rate

Valuation process starts with determination of risk-free rate for securities quoted at MSE, having in mind all relevant factors that affect emerging stock-markets, like low liquidity, small number of traded securities and short history of the market. Most risk and return models in finance start off with an asset that is defined as risk- free and use the expected return on that asset as the risk- free rate. In order to define an asset as risk -free it has to fulfill some requirements. In particular, an asset is risk- free if we know the expected returns on it with certainty, or when actual return is equal to the expected return. It means that first, there is no default risk for this type of security and second, there is no reinvestment risk (Damodaran, Damodaran on Valuation: Security Analysis for Investment and Corporate Finance , 2001). Macedonian Government as issuer of the Bonds for denationalization has to be viewed as a default free entity. When doing valuation, the risk- free rate should be the long term government bond rate that will be used as a discount rate. First option is to use risk-free rate of return on Treasury Security with ten years maturity issued by Government of the Republic of Macedonia. However, there are several reasons why this bond yield is not suitable for use as a discount rate. Macedonian government securities does not issue long-term denar-denominated state securities, T-Bonds are not issued regularly as well in amounts that can be planned in a advance, they are not zero-bonds which means that they have reinvestment risk, have low liquidity on capital market and they also have included country risk premium. It is important to emphasize that country risk premium can be added as separate element in changed CAPM equation. Due to the above mentioned reasons and in order to avoid calculation of country risk premium twice when use CAPM, we decide to use alternative model for risk-free rate calculation. Second option is to calculate risk-free rate from estimated 3% or 3.5% for ten-year euro-denominated bonds and adding the spread to risk-free interest rates, with the minimum estimated spreads, determined on the basis of expected annual inflation and Macedonia's credit rating (BBB, according to the Standard and Poors in 2009), being 3,3 percentage points for denar bills and 3,3 percentage points for long-term bonds. Third option is to use Svensson method for interest rate calculation, and based on that model to proceed with current German IDW method (the Institute of Public Auditors in Germany). In accordance with that methodology, the interest curve is established on the basis of a Svensson approximation (Ferenczi, 2006) and that fixed cash flows growing at a constant rate can be discounted using that interest curve.

We decide to use alternative methodology (Bloomberg 2009) and calculate risk-free rate by using 10-years Treasury Bonds denominated in euro issued by countries-members of European Union. We use yield-to-maturity (YTM) of these bonds with date of calculation, which has to represent forecasting of risk-free rates in the EU countries in the future. Due to the fact that this yields can be affected from volume of issue, we measure yields from forecasted GDP for these countries. Risk-free rate for MSE calculation presented in this paper was made in 2009 in order to make valuation of T-Bonds RMDEN09, so we have measured yields from forecasted GDP for countries members of EU for the entire 2009. We calculate single GDP – as a weighted average of YTM for 10-years Bonds denominated in euro, issued by European Governments. This process is presented in **Table 1**:**Error! Reference source not found.Table 1. Calculation of YTM on 10 – years EU Treasury Bonds (Date 30/09/2009)**

| Country | YTM 10y Bonds (%) | GDP (EUR bn) | GDP weight | W avg. YTM |
|---------|-------------------|--------------|------------|------------|
| 1 | 2 | 3 | 4 | 5 |
| Austria | 3,70 | 381,1 | 3,22% | 0,12% |
| Belgium | 3,72 | 436,7 | 3,69% | 0,14% |
| Finland | 3,59 | 232,1 | 1,96% | 0,07% |
| France | 3,54 | 2625 | 22,17% | 0,79% |
| Germany | 3,33 | 3107 | 26,25% | 0,87% |
| Greece | 4,46 | 354,3 | 2,99% | 0,13% |
| Ireland | 4,74 | 219,2 | 1,85% | 0,09% |
| Italy | 3,93 | 2073,3 | 17,51% | 0,69% |
| Holland | 3,57 | 785,5 | 6,64% | 0,24% |
| Portugal | 3,88 | 219,9 | 1,86% | 0,07% |
| Spain | 3,79 | 1403,7 | 11,86% | 0,45% |
| **Average** | | | **100%** | **3,65%** |

Description for above table: Second column presents YTM on 10- years T-Bonds issued from 11 countries members of EU. Next column (no.3) presents amounts of individual countries GDP in EUR. In column no.4 we calculate individual participation of countries GDP as percentage of Total EU countries GDP. In column 5 we calculate weighted average YTM - we multiple YTM on 10-years T-Bonds in % (column no. 2) and column no.4. Finally we sum all results from column no.5 and get weighted average 3,65%.

Valuation has to be done in real terms. This means that cash flows are estimated using real growth rates and without allowing for the growth from price inflations. It implies that discount rates used in these cases have to be real, not nominal risk-free rate (Damodaran, Investment Valuation: Second Edition, 2006). We have calculated weighted average YTM, that contains investors' expectations for future inflation rates in Euro–zone, so it is necessary to decrease nominal return in amount of such expected inflation rate, using Fisher Formula (Fisher, (1977) (1930)):

$$1 + r_n = (1 + r_r)(1 + i)$$

where

$r_n$ – nominal rate of return;

$r_r$ – real rate of return;

$i$ – inflation rate.

Expected inflation was calculated as geometric average of 10-years forecasting for Euro-zone (1,50%), based on European Central Bank forecasting. Using formula for nominal calculated weighted average for bonds denominated in euro, we calculate real YTM for these bonds (2,12%). We add to the real return, using Fisher formula, geometric average for 10-years expected rates of inflation in Macedonia (3,32%), in order to determine nominal, 10 – years, risk-free rate of return denominated in Macedonian national currency - Denars.

**Table 2. Calculation of Risk-Free Rate in Macedonia (Date 30/09/2009)**

| EU Treasury Bonds | Risk-free rate | Inflation EU | Real Risk-free rate | Inflation Macedonia | Risk-free rate Macedonia |
|---|---|---|---|---|---|
| 1 | 2 | 3 | 4 | 5 | 6 |
| EU T-Bonds | 3,65% | 1,50% | 2,12% | 3,32% | 5,49% |

Description for above table: Using Fisher formula we calculate real risk-free rate in EU. We add inflation rate calculated as 10-years geometric average and get risk-free rate in Macedonia.

Following this approach, 10-years denominated risk-free rate of return in Macedonia is calculated to be 5,49%. The entire above explained process is based on an assumption that purchasing power parity of Macedonian Denar and Euro will remain constant. Risk-free rate that will be used for valuation of Treasury bonds is 5,5%.

### 1.2. Valuation of Treasury Bond for Denationalization – RMDEN09

MSE started trading with Treasury Bonds RMDEN09 at 26.04.2010 in total amount of 30 million euros. This bond has same characteristics like previously issued Treasury Bonds for Denationalization (10 years maturity, the principal repaid over the life of the bond, i.e. - annuity payment of par value, 2% interest rate and first date of payment on 01.06.2011). It means that RMDEN09 and other Treasury Bonds for Denationalization have amortization schedule of principal and interest repayment. Analysis shows that this bond is attractive for investors due to possibility to be protected from the risk of possible depreciation/devaluation of Macedonian currency 'Denar' since bonds are denominated in euro, so they can provide protection of investors from foreign-exchange risks. This bond also gives the investors an opportunity to invest in risk-free instruments as there is evident lack of similar risk-free instruments on MSE, which constantly raises the demand for risk-free securities.

Valuation of Treasury Bond RMDEN09 is focused on determination of the intrinsic value of the bond. Price of a security in a competitive market should be the present value of the cash flows that investors will receive from owning it. In order to determine present value of future cash flows we will have to discount them with required rate of return (yield). Due to the fact that Treasury Bond is risk-free i.e. promised cash flow will be paid with certainty, we use the method of discounted cash flows and we discount bond's cash flows with required rate of return (risk-free rates) equal to 5,5% and determine the intrinsic value of RMDEN09 on 84,83. (Table 3).

**Table 3. RMDEN09 Amortization Plan and NPV Calculation (Nominal Value 100.000 EUR)**

| Year | Principal | Interest | Total Payment | Total amount of debt |
|---|---|---|---|---|
| 1 | 2 | 3 | 4 | 5 |
| 2011 | 10.000 | 2.000 | 12.000 | 90.000 |
| 2012 | 10.000 | 1.800 | 11.800 | 80.000 |
| 2013 | 10.000 | 1.600 | 11.600 | 70.000 |
| 2014 | 10.000 | 1.400 | 11.400 | 60.000 |
| 2015 | 10.000 | 1.200 | 11.200 | 50.000 |
| 2016 | 10.000 | 1.000 | 11.000 | 40.000 |
| 2017 | 10.000 | 800 | 10.800 | 30.000 |
| 2018 | 10.000 | 600 | 10.600 | 20.000 |
| 2019 | 10.000 | 400 | 10.400 | 10.000 |
| 2020 | 10.000 | 200 | 10.200 | - |
| NPV (5,5%) | 84.83 | | | |

In Table 4 we present average prices of Treasury Bonds for Denationalization in appropriate year.

**Table 4. Average bonds prices on MSE (2002-2011)**                                   **(in percentanges)**

| Bonds and Date of Issue | 2002 | 2003 | 2004 | 2005 | 2006 | 2007 | 2008 | 2009 | 2010 | 2011 |
|---|---|---|---|---|---|---|---|---|---|---|
| RMDEN01 25.06.2002 | 60% | 63.50% | 70% | 69% | 85% | 83,5% | 89,3% | 90% | 93% | 95,5% |
| RMDEN02 26.03.2003 |  | 50% | 65% | 68% | 82,6% | 84,1% | 88% | 89,7% | 92% | 98,3% |
| RMDEN03 01.03.2004 |  |  | 60% | 67,6% | 81,3% | 83,5% | 85% | 87% | 89% | 97,2% |
| RMDEN04 08.03.2005 |  |  |  | 60,9% | 80% | 82% | 84% | 85% | 89% | 96% |
| RMDEN05 15.03.2006 |  |  |  |  | 73.50% | 80% | 84,5% | 85% | 87% | 95,2% |
| RMDEN06 13.03.2007 |  |  |  |  |  | 78% | 85% | 81% | 86% | 94,5% |
| RMDEN07 26.08.2008 |  |  |  |  |  |  | 80% | 84% | 85% | 92% |
| RMDEN08 08.04.2009 |  |  |  |  |  |  |  | 79% | 81% | 92% |
| RMDEN09 26.04.2010 |  |  |  |  |  |  |  |  | 78,5% | 91% |
| RMDEN10 30.03.2011 |  |  |  |  |  |  |  |  |  | 90% |

Source:www.mse.com.mk

If we compare T-Bonds average market prices (in percentages of par value) with intrinsic value, we can see that bonds quoted and traded on MSE have lower market prices at the beginning of period of their issue compared with intrinsic value, which means that they were traded with discount. Their market prices have risen in following years and keep around intrinsic value and rise again in last period of bond maturity, which means that they are traded with premium. If we compare T-Bonds YTM calculated at MSE (rate that equal market price and present value of cash flows of the bonds), presented on Table No. 5, with our determined risk-free rate (5,5%), we can see that rates are equal only for RMDEN08. Three other T-Bonds (RMDEN07, RMDEN03 and RMDEN02) have 5% YTM which is near our calculated risk-free rate, while all others are below 5%.

**Table 5 YTM of T-Bonds ate MSE**

| RMDEN10 | RMDEN09 | RMDEN08 | RMDEN07 | RMDEN06 | RMDEN05 | RMDEN04 | RMDEN03 | RMDEN02 |
|---------|---------|---------|---------|---------|---------|---------|---------|---------|
| 3,77    | 4,33    | 5,5     | 5       | 4,67    | 4,86    | 4,87    | 5       | 5,07    |

Source: www.mse.com.mk

This mean that all T-Bonds at MSE (beside RMDEN08) were traded with premium. A deeper analysis of bond price fluctuations indicates leads to a conclusion that discounted price of the bonds in first years from issuing are due to the higher volume of traded Treasure Bonds, i.e. bigger supply on the market when bond holders are trying to sell their bond portfolios and get required return faster. As a result of strong demand for risk-free instruments especially from institutional investors (pension funds, insurance companies and investment funds) that are obliged by Law to keep significant part of their portfolios in risk-free instruments, this situation increases demand in the next period and provokes rise of the bond prices and they were traded with premium.

Although Treasury Bonds for Denationalization promised just 2% interest rate, they provide higher yield to maturity compared with similar investment opportunities on financial markets in the Republic of Macedonia, due to possibility for reinvestment. Higher demand for the bonds increases bond market prices.

## 2. Volatility of the Bonds: Empirical Test of the Duration, Modified Duration and Convexity of the Treasury Bonds on MSE

There is difference between nominal maturity and time of effective return of initial investment in bonds. In order to see the difference we will calculate duration of the Treasury bonds on MSE. Duration is weighted-average time needed for effective return of investment in bonds. We use as weighted average- present value of interest and principle payment and we divide all of them with bond price. Duration is given by following equation (Arnold, 2008), and it is measured in years:

$$D = \frac{\sum t\, PVCF}{P} \tag{1}$$

Modified duration is defined as:

$$D^* = D/(1 + y) \tag{2}$$

Bonds with shorter duration are less sensitive on interest-rate changes compared with bonds with longer duration. Duration is expressed as calculated average maturity of the bond, where we use discounted cash flows for each period (DeMarzo, 2008). Our calculation of Macaulay duration shows influence of different bonds maturity on duration. Using duration we can quantify bond sensitivity on interest rate change, maturity and bond price as given by the following equation:

$$\frac{\Delta P}{P} = -D^* \Delta y \tag{3}$$

Key bond-interest rate relationships are that bond prices are inversely related to changes in market interest rates. This means that all else equal, long-term bonds are more sensitive to interest rate changes than short - term bonds. All else equal, low-coupon bonds are more sensitive to interest rate changes than high-coupon bonds. Bond prices are inversely related to the market rate of interest. Bond convexity can be explained as - all else equal, the higher duration (longer time to maturity or lower coupon payment), the more convexity will be and all else equal, the bigger the change in interest change, the more convexity will be. Convexity is given by following equation:

$$C = \frac{\sum (t^2 + t)\, PVCF}{P(1 + y)^2} \tag{4}$$

Duration and convexity can be used to estimate the sensitivity of bond price on changes in interest rate:

$$\frac{\Delta P}{P} = -D^* \Delta y + 0.5 C (\Delta y)^2 \tag{5}$$

We start our calculation of Duration of RMDEN10. We use interest rate i=2%, YTM (y) is 3,77%, maturity (t) is 10 years and nominal value (M) is 100. Duration of RMDEN10 is 5,04 years calculated using Eq. (1). Modified duration is calculated using Eq. (2) to be $5,04/(1 + 3,77\%)=$

4,865159 years. Convexity is calculated from Eq. (4) and for RMDEN10 is equal to 3539,54*0.01014147= 35,89614. If interest rate *y* rises to 4% which is an increase of Δy = 0,23%, then the price of the bond (see Eq. (3)) will drop to 90,54585 which is a decrease of -1,02466%:

$$\frac{\Delta P}{P} = -D^*\Delta y = -4,865 * 0,0023 = -0,01119$$

$$\Delta P = 91,57*(-0,01119) = -1,02466\%$$

$$P = 91,57 - 1,02466 = 90,54585$$

We can also confirm that there is small difference between forecasting of bond price change with duration and discounting of bond cash flow directly with 4% discount rate. This difference is minimal, it means that by discounting directly all cash flows with 4% yield to maturity we get almost same result (bond price 90,55%). We can derive conclusion that duration can be used with great certainty for forecasting of bond prices change for the bonds with 10 years maturity.

Next we analyze the possibility to use convexity for forecasting of bond price change. Making the same assumptions that market interest rates increase for 0,23%, bond price will decrease for 1,02308%, and new bond price will be 90,554% (see Eq.(5)):

$$\frac{\Delta P}{P} = -D^*\Delta y + 0.5C(\Delta y)^2 = -4,865159 * (0,0023) + 0.5 * 35,89614(0,0023)2 = -0,01109$$

$$\Delta P = 91,57*(-0,01109) = -1,01596$$

$$P = 91,57 - 1,01596 = 90,55404$$

We can conclude that convexity effect is relatively small for analyzed bond, and due to the fact that there is no difference between forecasting of bond price change with convexity and discounting of bond cash flow directly with 4% discount rate, convexity can be used as accurate measure for bonds price forecasting.

Figure 1 gives the sensitivity of bond value $P$ (RMDEN10) on changes in the interest rate $y$. Interest rate $y$ is given on the $x$-axis, while $\Delta P$ is given on the $y$ axis. First, $\Delta P$ is calculated using the correct formula:

$$\Delta P = P(y + \Delta y) - P(y) \qquad (6)$$

and then it is estimated using duration (see Eq.(3)) and convexity (see Eq.(5). In a wide range of values for $y$ between 2.5% and 5%, true change in bond value (circles) can be closely estimated using Duration (squares), and using Duration and Convexity (triangles). Dotted line with triangles almost completely overlaps and hides the solid line with circles since the true change in bond value can be estimated with high precision using Duration and Convexity.

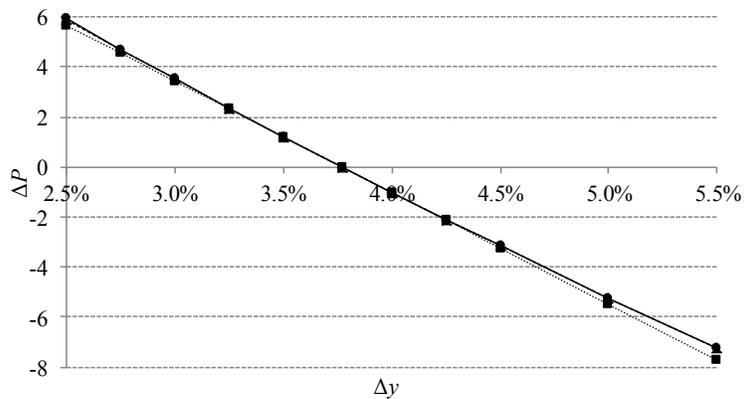

**Figure 1. Sensitivity of bond value $P$ on changes in the interest rate $y$.**

Figure 2 depicts how correct the estimation is. Dependence of estimation errors $\frac{P_D - P}{P}$ and $\frac{P_C - P}{P}$ on the change $\Delta y$ in the interest rate is shown. Clearly, estimation of bond value $P$ using both Duration and Convexity significantly reduces the estimation error compared to the case when only Duration is used.

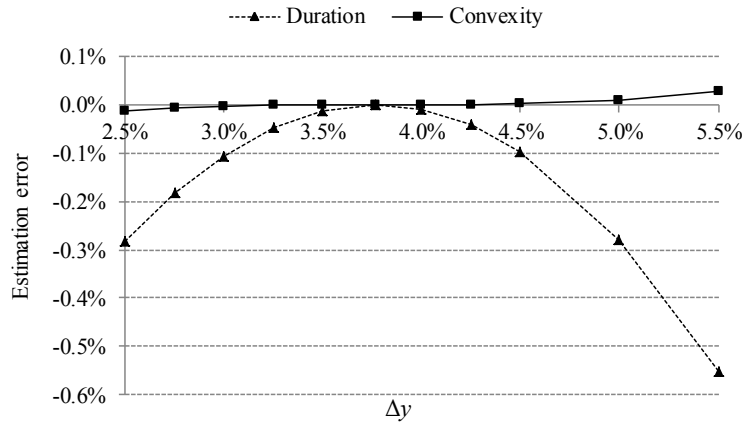

**Figure 2. Dependence of estimation error on changes in the interest rate *y*.**

Table 6 gives the data from Figure 1 and Figure 2 in a tabular form. Figure 1 visualizes rows $\Delta P$, $\Delta P_D$ and $\Delta P_C$, while Figure 2 visualizes last two rows from Table 6.

**Table 6. Estimation of bond value *P* using Duration and Convexity.**

| Y | 2.50% | 2.75% | 3.00% | 3.25% | 3.50% | 3.77% | 4.00% | 4.25% | 4.50% | 5.00% | 5.50% |
|---|---|---|---|---|---|---|---|---|---|---|---|
| P | 97.504 | 96.291 | 95.101 | 93.932 | 92.785 | 91.571 | 90.554 | 89.469 | 88.404 | 86.330 | 84.330 |
| $P_D$ | 97.228 | 96.115 | 95.001 | 93.887 | 92.773 | 91.571 | 90.546 | 89.432 | 88.318 | 86.091 | 83.863 |
| $P_C$ | 97.494 | 96.286 | 95.098 | 93.932 | 92.785 | 91.571 | 90.555 | 89.470 | 88.406 | 86.339 | 84.355 |
| $\Delta P$ | 5.934 | 4.721 | 3.530 | 2.362 | 1.215 | 0.000 | -1.016 | -2.101 | -3.167 | -5.240 | -7.240 |
| $\Delta P_D$ | 5.658 | 4.544 | 3.430 | 2.317 | 1.203 | 0.000 | -1.025 | -2.138 | -3.252 | -5.480 | -7.707 |
| $\Delta P_C$ | 5.923 | 4.715 | 3.528 | 2.361 | 1.215 | 0.000 | -1.016 | -2.101 | -3.165 | -5.231 | -7.215 |
| $(P_D-P)/P$ | -0.2828% | -0.1833% | -0.1049% | -0.0481% | -0.0130% | 0.0000% | -0.0095% | -0.0417% | -0.0969% | -0.2776% | -0.5539% |
| $(P_C-P)/P$ | -0.0109% | -0.0057% | -0.0025% | -0.0008% | -0.0001% | 0.0000% | -0.0001% | -0.0006% | -0.0022% | -0.0105% | -0.0294% |

Next calculation is Duration of RMDEN09. We use interest rate i=2%, YTM (y) = 4,33%, maturity (t) = 9 years, and nominal value (M) = 90. RMDEN09 Duration is 4.598 years. Modified Duration is 4,4075, and Convexity is 29,67973. If we assume that market interest rates (YTM) decreases from 4,33% to 4%, which means decrease of 0,33%, then using duration we can calculate bond price change (increase) for 1,3088%:

$$\frac{\Delta P}{P} = -D^*\Delta y = 4,4075 * 0,0033 = 0,014545$$

80.98552/90*0,014545=1,308806%

New bond price will be 80.98552/90+1,3088%=91,2927%

If we discount bond cash flows with 4% YTM, then we get bond price of 91.30%, which is a small deviation from duration. If we use same assumptions for yield decrease for 0,33% and forecast bond prices change with convexity, we get bond prices change for 1,323% and new bond price of 91,307, which means small deviation of the bond:

$$\frac{\Delta P}{P} = -D^*\Delta y + 0.5C(\Delta y)^2 = 4.407 * (-0.0033) + 0.5 * 29.679 * (0.0033)2 = 1,3233\%$$

$$80.98552/90+1,3233\%=91,307\%$$

Figure 1 gives the sensitivity of RMDEN09 bond value *P* on changes in the interest rate *y*, in a wide range of values for *y* between 3.25% and 5.5%. True change in bond value (circles) can be closely estimated using Duration (squares), and using Duration and Convexity (triangles).

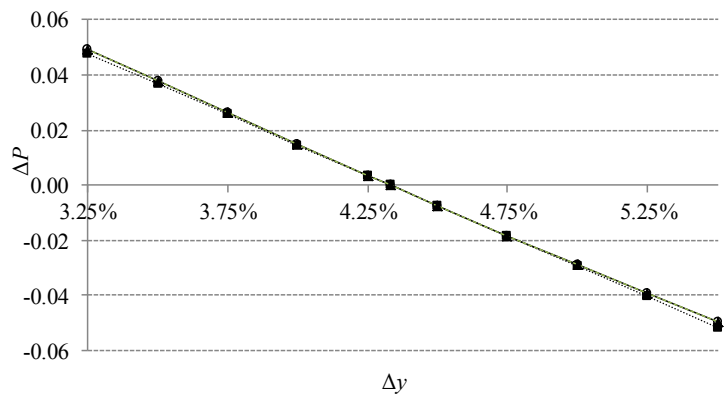

**Figure 3 Sensitivity of bond value *P* on changes in the interest rate *y*.**

Figure 24 depicts how correct the estimation is. We draw the same conclusion as in Figure 2. Namely, estimation of bond value *P* using both Duration and Convexity significantly reduces the estimation error compared to the case when only Duration is used.

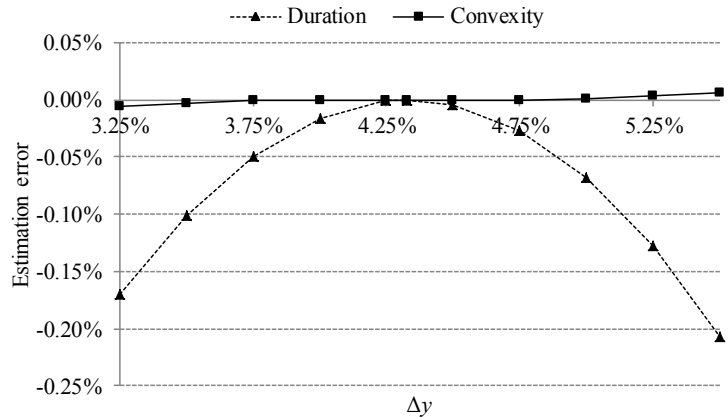

**Figure 4 Dependence of estimation error on changes in the interest rate *y*.**

We have calculated Duration, Modified duration and Convexity for all other T- Bonds traded at MSE. They are presented in Table 11.

**Table 7 Duration, Modified duration and Convexity of Treasury Bonds on MSE**

| BONDS | D | Dmod | Conv |
|---|---|---|---|
| RMDEN10 | 5,04 | 4,86 | 35,89 |
| RMDEN09 | 4,59 | 4,40 | 29,67 |
| RMDEN08 | 4,12 | 3,90 | 23,62 |
| RMDEN07 | 3,73 | 3,55 | 19,26 |
| RMDEN06 | 3,31 | 3,16 | 15,68 |
| RMDEN05 | 2,86 | 2,73 | 11,89 |
| RMDEN04 | 2,41 | 2,30 | 8,64 |
| RMDEN03 | 1,95 | 1,86 | 5,84 |
| RMDEN02 | 1,48 | 1,411 | 3,56 |

Description of above table: Calculated Duration, Modified Duration and Convexity for all Treasury Bonds on MSE presented in one table.

      We conclude that when forecasting bonds price changes, deviations are smaller for bonds with shorter maturity. This means that duration is good approximation and that Treasury Bonds traded at MSE have lower convexity. This also confirms the fact that convexity is better measure for bond price changes

prediction on MSE compared with duration of the bonds. Duration determines bonds sensitivity on interest rate changes and shows approximately the time in which risk of price changes offset the reinvestment risk. Holding bonds in period of duration immunize investors from interest rate changes.

The empirical results also provide support of the existence of a non-linear relationship between interest rate risk and prices of the T- Bonds for Denationalization on MSE. If we analyses T-Bonds price changes at MSE we can conclude that T- Bonds for denationalization on MSE that are not issued regularly and in equal series do not react on interest rate changes. As previously elaborated there are no other risk-free securities at MSE as well financial derivatives that can be used for hedging or for portfolio optimization of institutional investors. This keep demand for T-Bonds higher and make them not sensitive on interest rate risk.

### 3. Summary and Concluding Remarks

This study presents practical valuation of T-Bonds at MSE in order to determine how standard valuation models fit in case of T- Bonds that are traded on MSE and to verify whether they offer reliable results compared with average bonds prices at MSE. We compare T-Bonds average market prices (in percentages of par value) with intrinsic value of the bonds, and conclude that bonds quoted and traded on MSE have lower market prices at the beginning of period of their issue compared with intrinsic value, which means that they were traded with discount. Their market prices have risen in following years and keep around intrinsic value and rise again in last period of bond maturity, which means that they are traded with small premium. All T-Bonds quoted at MSE (beside RMDEN08) were traded with premium. A deeper analysis of bond price fluctuations leads to a conclusion that discounted price of the bonds in first years from issuing are due to the higher volume of traded T- Bonds, i.e. bigger supply on the market when bond holders are trying to sell their bond portfolios and get required return faster. As a result of strong demand for risk-free instruments especially from institutional investors (pension funds, insurance companies and investment funds) that are obliged by Law to keep significant part of their portfolios in risk-free instruments, this situation increases demand in the next period and provokes rise of the bond prices and they were traded with premium. This results suggest that for bond valuation at MSE, beside intrinsic value calculation it is necessary to take into consideration YTM and Total Return of the bonds with reinvestment.

We also suggest alternative methodology for risk-free rate calculation at MSE, that use weighted average YTM of 10-years Treasury Bonds denominated in euro issued by countries-members of European Union. We use yield-to-maturity (YTM) of these bonds with date of calculation, which has to represent forecasting of risk-free rates in the EU countries in the future. Due to the fact that this yields can be affected from volume of issue, we measure yields from forecasted GDP for these countries for the entire 2009. We calculate single GDP – as a weighted average of YTM for 10-years Bonds denominated in euro, issued by European Governments. Using Fisher formula we eliminate inflation in EU zone and add expected inflation in the Republic of Macedonia. This leads to final calculation of risk-free rate in the Republic of Macedonia. This methodology eliminates shortages of using directly Macedonian T-Bonds long-term bond yield as a discount rate. Macedonian government securities are not issued regularly as well in amounts that can be planned in a advance, they are not zero-bonds which means that they have reinvestment risk, have low liquidity on capital market and they also have included country risk premium. It is important to emphasize that country risk premium can be added as separate element in changed CAPM equation. Due to the above mentioned reasons and in order to avoid calculation of country risk premium twice when use CAPM, we offer alternative model for risk-free rate calculation.

We also make empirical test of duration, modified duration and convexity of the T-bonds at MSE in order to determine sensitivity of bonds prices on interest rate changes. Key relationship between bond prices and interest rate is that bond prices are inversely related to changes in market interest rates, calculating duration, modified duration and convexity we test the sensitivity of T- Bonds for Denationalization on MSE on interest rate changes and determine the measure that is better for bond prices forecasting. We analyze annual data that covered the 2001-2011 sample period. The empirical results provide evidence that convexity is more accurate measure as approximation of bond prices changes than duration.

Final conclusion of this study is that T-Bonds traded at MSE are not sensitive on interest rate changes due to institutional investors' permanent higher demand and at the same time market limited offer of risk-free instruments.